\documentclass[letterpaper,12pt]{article}  

\pdfoutput=1 
\usepackage{subfigure}
\usepackage{amsfonts}
\usepackage[pdftex]{graphicx}
\usepackage{amsmath,amssymb,amsthm}
\usepackage{natbib}
\usepackage{hyperref}
\usepackage{bm}
\usepackage{caption}
\usepackage{subcaption}
\usepackage{adjustbox,lipsum}
\usepackage{fancyhdr}
\usepackage{sectsty}
\usepackage{stmaryrd}
\usepackage{setspace}   
\usepackage{longtable}                   
\usepackage{booktabs}    
\usepackage{pdfsync}        
\usepackage[backgroundcolor=blue!40,linecolor=blue!40]{todonotes}
\usepackage{fancybox}
\usepackage{appendix}

\definecolor{cornellred}{RGB}{179,27,27} 
\definecolor{cornellblue}{RGB}{00,00,170}
\definecolor{cornellgrey}{RGB}{96,94,92}
 
\theoremstyle{myplain}
\newtheorem{theorem}{Theorem}[section]
\newtheorem{corollary}{Corollary}[section]

\newtheorem{lemma}{Lemma}[section]

\newtheorem{definition}{Definition}[section]
\newtheorem{example}{Example}[section]

\theoremstyle{myremark}

\renewcommand{\cite}{\citet}

\usetikzlibrary{calc}
\def\centerarc[#1](#2)(#3:#4:#5){ \draw[#1] ($(#2)+({#5*cos(#3)},{#5*sin(#3)})$) arc (#3:#4:#5);}

\hypersetup{colorlinks=true, linkcolor=blue, citecolor=blue}                          
\numberwithin{equation}{section}

\begin{document}
\title{A Better Test of Choice Overload\thanks{We recognize advice and input from Benjamin Scheibehenne, the members of the Cognition and Decision Laboratory at Columbia University, and the Columbia Experimental Laboratory in the Social Sciences. We thank audiences at Berkeley Haas, Harvard, Princeton, Stanford, UCLA, the University of Queensland, the Pan-Asian Theory Seminar, the MiddExLab Virtual Seminar, the Zurich Workshop on Economics and Psychology, the 2023 Bounded Rationality in Choice Conference, the 2023 Econometric Society European Summer Meeting, and the 2023 Econometric Society North American Summer Meeting for feedback and Brenda Quesada Prallon as well as Estelle d'Alessio for careful readings of the manuscript. We gratefully acknowledge financial support from Columbia's Experimental Laboratory in the Social Sciences and through NSF grant SES-1824375.}}
\author{Mark Dean\thanks{Department of Economics, Columbia University. Email mark.dean@columbia.edu.}\and Dilip Ravindran\thanks{Humboldt University Berlin. Email: dilip.ravindran@hu-berlin.de.}\and J\"{o}rg Stoye\thanks{Department of Economics, Cornell University. Email stoye@cornell.edu.} }
\date{\today \\}

\maketitle
\begin{abstract}
Choice overload -- in which larger choice sets are detrimental to a chooser's well-being -- is potentially of great importance in the design of economic policy. Yet current evidence on its prevalence is inconclusive. We argue that existing tests are likely to be underpowered and hence that choice overload may occur more often than the literature suggests. We propose more powerful tests based on richer data and characterization theorems based on specific models, including random utility. These new approaches come with significant econometric challenges, which we show how to address. We apply our tests to new experimental data and find strong evidence of choice overload that would likely be missed using current approaches. 
\end{abstract}

\bigskip

\section{Introduction}
Standard economic reasoning asserts that increasing the set of options cannot make the consumer worse off. Yet, starting with the pioneering study of \citet[][see also \citet{reibstein1975number}]{iyengar2000choice}, a growing body of work on \emph{choice overload} --  broadly speaking, the idea that people make worse choices in larger choice sets -- has called this assumption into question. If true, choice overload would have important normative and positive implications for economics. It is inconsistent with utility maximization and many classic behavioral models. It also calls into question policy recommendations based on giving consumers as large a range of options as possible from which to choose.

Unfortunately, current research on choice overload is inconclusive. Some direct replications of previous experiments failed \citep{scheibehenne2008effect, greifeneder2010less}. One recent meta-analysis concludes that the mean measured choice overload effect is zero \citep{scheibehenne2010can}, and one \citep{chernev2015choice} concludes that its relevance may heavily depend on context. At the very least, there remains debate as to whether choice overload is a widespread and robust phenomenon. 

In this paper, we argue that existing studies are likely to underestimate the extent of choice overload. A standard experimental paradigm for identifying overload is to compare the frequency of choosing a default option in small and large choice sets -- for example, sticking with the default of not buying a jam in \citeauthor{iyengar2000choice}'s (\citeyear[][IL henceforth]{iyengar2000choice}) classic study.\footnote{While other measures of overload are used, such as ex post satisfaction and regret, \citet{chernev2015choice}  shows them to be highly correlated.} A data set is said to exhibit choice overload if and only if the default is \textit{more} likely to be chosen in a larger choice set than it is in the  smaller choice set. Yet intuitively, under the hypothesis of utility maximization one would expect the default to be chosen \textit{much less} in larger choice sets as the number of potentially attractive options grows. This lack of power may explain inconclusive results from existing tests. 

Three examples may further clarify this point.
\begin{example} \label {pref_het} \textbf{Preference heterogeneity}
Consider a simple model of the classic ``jam'' experiment: Subjects have a probability $p$ of liking any particular jam better than ``no jam"; for sake of argument, suppose this preference is independent across subjects and jams (we will not maintain this assumption for the rest of the paper). IL found the probability of buying a jam from a set of six to be $12\%$. In this model, this would mean $p$ is approximately $2\%$, and the implied probability of buying a jam from the 24-jam set would be $1-.88^{24/6}\approx 40\%$. Under the admittedly strong independence assumption, it would therefore require extreme choice overload to push active choice below $12\%$ in the larger choice set -- and yet this is the benchmark typically used to identify the effect. 
\end{example}
\begin{example} \label{item_het} \textbf{Item heterogeneity}
Consider another variant of the IL example: Of the 24 jams IL use, 23 are variants of gooseberry jam, and one is strawberry jam. 10\% of people like gooseberry jam, while everyone likes strawberry jam. This means that, absent choice overload, the probability of buying jam from the 24 jam set would be 100\%. For a smaller choice set, it would be 100\% if the strawberry jam were included or 10\% otherwise. If the smaller set of jams selected by the researched happened not to include strawberry jam, then choice overload would have to be extremely strong to put active choice below the 10\% benchmark. 
\end{example}
\begin{example} \label{ex:min} \textbf{Tighter bounds}
Let the choice universe be $X=\{a,b,c,d\}$, where $d$ stands for default. Suppose $a,b,c$ are individually chosen across a population with probability $20\%$ from choice sets $\{a,d\},\{b,d\},\{c,d\}$. Suppose also that the probability of non-default choice is $40\%$ from any of $\{a,b,d\},\{a,c,d\},\{b,c,d\}$. Assuming no indifferences, these probabilities \emph{can} be reconciled with a Random Utility Model (RUM, precisely defined later); however, they imply that the preferences $a \succ d$, $b \succ d$, and $c \succ d$ occur in disjoint subsets of the population. Thus, non-default choice probability from $X$ must be at least $60\%$, even though the largest such probability observed in smaller sets is ``only'' $40\%$.\end{example}

We introduce new tests for choice overload for a data set in which default versus nondefault choice is observed for a collection of subsets from a grand set $X$ (always including the same default option). We define choice overload as a situation in which the default is chosen too often in $X$ relative to some benchmark. We first equate choice overload with violations of monotonicity -- that is, the probability of choosing the default option increases as choice sets expand. This is the approach previously used in the literature, though the structure of our data still allows for an improved test.

We next extend our definition to identify choice overload \textit{relative to some model}. Our primary example is RUM -- that is, choices can be explained by a (choice set independent) probability distribution over a set of utility functions. As example \ref{ex:min} illustrates, consistency with RUM implies monotonicity, but not vice versa, so the latter definition provides a more sensitive test for choice overload.\footnote{In general, it is well known that stochastic monotonicity is necessary but not sufficient for random utility. Example \ref{ex:min} clarifies that coarsening the domain of observations to active versus passive choice does not change this.} However, one could use the same definition to identify choice overload relative to other models, as we illustrate using the Rational Shortlist Method (RSM) of \citet{manzini2007sequentially}.

If the data consists only of choice probabilities from $X$ and a single subset $A$, then bounds based on monotonicity and RUM coincide and reduce to the current best practice. We therefore propose that data should be collected from multiple subsets of $X$. 
In this richer data, monotonicity implies that choice probability of the default from $X$ is less frequent than from any other choice set $A$. We call this the \textit{Min bound}. While easy to define, testing this restriction is subtle because it is a test of multiple hypotheses; we propose an approach that takes inspiration from closely related problems in the literature on moment inequalities. 

We next observe that our RUM-based definition of choice overload implies tighter bounds. We lean on classic work by \citet{McFadden1991} to characterize the highest default choice probability in the large set that is consistent with the data from subsets and RUM, which we call the \textit{RUM bound}. To implement this condition in practice, we adapt \citeauthor{KS18}'s (\citeyear{KS18}, see also \citet{RUM_compute}) nonparametric test of RUM.

We apply all tests to a novel experimental data set of choices from subsets of size $2$ and $3$ of $12$ different choice objects plus a default. The objects are verbally described sums of numbers as used in \citet{caplin2011search}; see Figure \ref{fig:1} for a preview.  Subjects were asked to chose from $10$ such sets, one of which was the default plus the entire set of $12$ other options (henceforth called the \textit{grand set}).

Traditional approaches would be unlikely to find evidence of choice overload in our data set: The average default choice in small sets is higher than in the grand set, and only 15 of the 78 small sets have a default choice which is significantly lower than that of the grand set.  In contrast, even the most brute force of our proposed tests -- a potentially very conservative but finite-sample valid implementation of the Min bound -- does detect it.

Our test based on RUM provides an unexpected additional insight: While RUM is indeed rejected on the whole data set due to choice overload, it is also rejected if we remove data from choices in the grand set. This is because subjects are more likely to choose the default option in choice sets of size $3$ than RUM would predict given their data from size $2$ sets. Thus, our results indicate that choice overload type effects can start in choice sets which are much smaller than was previously demonstrated.

We conclude this section with four important clarifications. First, our notion of choice overload is ``thin'' or behavioral. That is, we think of choice overload as a specific feature of choice behavior: that the default is chosen too often in larger sets. We do not commit to a theory of what is going on ``under the hood.'' For example, the increase in default choice set in larger sets observed in our data is plausibly explained by rational inattention. If true, we would describe this a a case of rational inattention \textit{causing} choice overload, rather than an alternative description of what is going on. We naturally think that ``deep'' theories of the mechanisms generating choice overload are important and we hope that our work aids future researchers in uncovering them; however, our main aim in this paper is not to substantively advance that inquiry but to provide tools with which to identify choice overload.    

Second, we do not see the main interest in our experiment to be whether or not choice overload exists in this particular environment. Rather, the point is whether or not we can \textit{detect} it. Because of this, we deliberately picked an (arguably somewhat artificial) environment we believed likely to cause choice overload. We then show that standard approaches would likely not have detected it, but our approaches can. It is this final point we see as the key takeaway from the experiment. 

Third, it is true that in our data set the additional technicalities and machinery needed for the test against RUM are unnecessary - choice overload is observable using the weaker monotonicity bound. However, that does not mean that in other data sets the additional power of testing against RUM would not be valuable. Arguably we did too good a job of designing an environment to generate choice overload, so our more technical approaches were not needed. This may not be true in other cases -- particularly if preference heterogeneity is important -- as we discuss further below.  

Fourth, we note that our tests can be applied to any data -- not just experimental -- where a large sample of default vs non-default choice is observed from a choice set and multiple subsets. For example, data on health insurance plan choices, such as that used in \cite{abaluck2023less},  often has variation in the specific plans offered as well as the number offered which could be used for our analysis. We leave such applications of our tests for future research.

\section{Literature Review}\label{Lit review} 

Two recent meta-analyses \citep{chernev2015choice,scheibehenne2010can} report results from 99 experiments in 53 papers and 63 experiments in 50 papers, respectively. We refer to these reviews and are content to just make two points.

First, very few current studies have generated the data needed to perform our tests. Using the aforementioned meta-analyses and Google Scholar, we identified 32 studies from 19 papers that use default choice as a measure of choice overload. Of these, 20 ask subjects to make choices from a single subset of the grand choice set. These studies can do no better than to compare the default choice probability in the small and large choice set. The remaining 12 studies ask subjects to make choices from multiple subsets of the large choice set, but only collect a small number of choices from each. As a result, they instead compare the average default choice across all small choice sets to that in the larger set, a measure which is necessary but not sufficient for the aforementioned Min bound, which in turn is necessary but not sufficient for consistency with RUM. While applying our approach to data from these previous studies is an interesting avenue for future research, the small number of observations in each small set make our tests hard to implement.

Although not explicitly designed to test for choice overload, the experiments of \citet{aguiar2023random} contain the type of data needed to perform our test. Choices were observed from all subsets of a grand set of six lotteries, with a default that is always available. The authors report no evidence of choice overload, and our tests confirm this result. This may be due to the fact that the default alternative in their experiment was chosen to be obviously worse than the other available alternatives. 

The second point is that the veracity and scope of choice overload is far from established. Some direct replications have failed \citep{scheibehenne2008effect}. One meta-analysis \citep{scheibehenne2010can} finds the mean effect of set size on measures of choice overload to be zero, but notes a high variance. A more recent analysis \citep{chernev2015choice} identifies four variables which can increase the incidence of choice overload: decision difficulty (for example due to time constraints), choice set complexity (for example due to hard-to-compare alternatives), preference uncertainty (for example because the decision maker is unsure how to aggregate their preferences across many dimensions), and decision goal (for example because the decision maker is not really committed to making a purchase). 

Finally, our testing problem is related to that studied in \cite{kono2023axiomatization}.\footnote{A recent paper by \citet{turansick25} is more distantly related. He considers characterizations of RUM when not all choice problems are observable, but for those that are all probabilities are observed.} We both test RUM when not all choice probabilities from a choice set are observed. While also taking inspiration from \cite{McFadden1991}, \citeauthor{kono2023axiomatization} identify a different set of conditions using the Block-Marschak polynomials. This approach is elegant and may have computational benefits, but there are some reasons why our approach is more directly applicable to the task at hand. First, while \cite{KS18} (and we) use RUMs as motivation, they (and we) really just test whether vectorized choice probabilities lie in a certain polyhedral cone. In application to RUM, the vertices of this cone are defined by ``choice types'' that are conventional utility maximizers, but one can analogously use this approach to test the ``random utility extension'' of other models of behavior (we do this with RSM, and generalizations of RUM that capture choice overloaded behavior). In addition, \citet{kono2023axiomatization} assume that the collection of observable choice sets is closed under set expansion; we do not need this and it does not hold in our data.

\section{Theory}

We now present the theoretical underpinnings of our test. We initially assume that we can perfectly observe default choice probabilities for each choice set; in a second step, we develop the econometric tools required because our actual data are finite samples. Additional computational considerations that proved unnecessary for our experiment are relegated to supplementary materials.

\subsection{Population-Level Analysis}\label{sec:population}

Let $X$ be a finite set of alternatives and $d$ a default alternative contained in $X$. Let $\mathcal{D}\subset 2^{X}/\emptyset $ be a collection of choice sets, all of which contain $d$; in most of what follows, and in our experimental data, $\mathcal{D}$ will contain $X$. 

Suppose initially that we observe a function $p_d:\mathcal{D}\rightarrow [0,1]$, where $p_d(A)$ is the probability of choosing the default $d$ from choice set $A$. We assume that we observe the population probability with which the default is chosen in each choice set, but not that we can track individuals across choice problems. This makes our approach applicable to many between-subject data sets. Crucially, we also assume that we observe only the probability with which the default was chosen in each choice set, not the probability with which specific non-default options are chosen. This is consistent with our desire to focus on choice overload effects; the limited data only allows us to detect violations of utility maximization that occur to due too much or too little default choice from a set, and the tests we will define will identify the former violations as `choice overload'.

We next consider two possible definitions of choice overload. The first one equates it with violations of choice monotonicity. 

\begin{definition}
Probabilities  $p_d$ satisfy monotonicity if, for any $A,B \in \mathcal{D}$ such that $A \subset B$, 
\begin{equation*}
p_{d}(A) \geq p_{d}(B)
\end{equation*}
\end{definition}
The canonical choice overload experiment, in which $\mathcal{D}=\{A,X\}$, tests this condition.
More generally, one can define
\begin{equation*}
p_{d}^{min}(X)=\min_{A\in \mathcal{D}\setminus X}p_{d}(A)
\end{equation*}
as the smallest observed probability of choosing $d$ in any set other than $X$. Monotonicity is violated iff $p_{d}(X)>p_{d}^{min}(X)$. We therefore refer to $p_{d}^{min}(X)$ as the Min bound. We will say that data which violates this condition exhibits choice overload with respect to the Min bound. Such data is inconsistent with a number of models --- most obviously RUM, but also models of reference dependent preferences such as \citet{tversky1991loss}.\footnote{This is also true for the stochastic consideration set model of \citet{manzini2014stochastic}, a special case of RUM. Other more general models of stochastic consideration allow for choice overload type effects -- see, for example, \citet{cattaneo2020random, cattaneo2021attention}.}

A second approach is to define choice overload as violation of a specific model.  The basic idea is that one would declare a data set to exhibit choice overload if the default is chosen too often in some large set $A$, given the choice probabilities from the smaller sets and the constraints imposed by the model of interest. Specifically, for any $A \subseteq X$, we can define a maximal bound on the choice of default using the default choice probabilities from the subsets of $A$ and consistency with the model, and define choice overload as default choice in excess of that bound. The basic idea here goes back to \citet{Varian82,Varian83}: counterfactual choice behavior is in the predictive bounds if, and only if, that choice behavior and previously observed ones (in our case, behavior on small choice sets) are jointly rationalizable. Formally:

\begin{definition}\label{def:p_mod}
Define
\begin{equation*}
p_d^{MOD}(A)=\sup x\in [0,1]
\end{equation*} 
subject to probabilities
\begin{equation*}
\tilde{p}_{d}(\tilde{A})=\left\{ 
\begin{array}{c}
x\text{ if }\tilde{A}=A  \\ 
p_d(\tilde{A})\text{ otherwise}
\end{array}
\right. 
\end{equation*}
on choice sets $\mathcal{D}_A \equiv \{\tilde{A} \in \mathcal{D}:\tilde{A} \subseteq A\}$ being jointly consistent with the model MOD. 
\end{definition}

We call a data set \textit{stochastically rationalizable} by model MOD if it could have been generated by that model. Then we can think of it as revealing choice overload if it would be stochastically rationalizable except that the probability of default choice in larger choice sets is too large.  One could implement this either by testing for the validity of MOD both including or excluding data from $X$, or by computing counterfactual bounds on $p_d(X)$ as the set of all default choice probabilities on $X$ that would be rationalizable jointly with the other observed probabilities.

The most obvious candidate model for us to consider is RUM, for three reasons; it represents the benchmark extension of rational choice to a stochastic setting; it encompasses most of the most popular models of stochastic choice (such as logit and probit); and it provides tighter bounds for choice overload than monotonicity alone. For the majority of the paper we will make use of a definition based on RUM, though below we also consider the implications of bounds based on the Rational Shortlist Method of \citet{manzini2007sequentially} and \citet{apesteguia2013choice}.


\begin{definition}
Probabilities $p_d$ are consistent with RUM if there exist a finite collection $\mathcal{U}$ of one-to-one utility functions on $X$ and a probability distribution $\rho \in \Delta(\mathcal{U})$ such that, for every $A \in \mathcal{D}$,
\begin{equation*}
p_d(A)=\sum_{u\in \mathcal{U}\mid d=\arg \max u(A)}\rho (u)
\end{equation*} 
\end{definition}

For most of this paper, we say that a data set exhibits choice overload according to the RUM bound if $p_{d}(X)>p^{RUM}_{d}(X)$, where $p^{RUM}_{d}(X)$ is an instance of Definition \ref{def:p_mod}. However, the definition of $p_d^{RUM}(\cdot)$ allows for choice overload to ``kick in" for smaller choice sets and we will consider that later.

The Min and RUM bounds speak to different reasons why observations from a single small choice set might not effectively identify choice overload. The first is heterogeneity in the quality of available items, as illustrated in example \ref{item_het}. Consider a grand choice set that consists of a number of not very appealing jam flavors and one extremely appealing flavor (strawberry, say). Absent any choice overload, we would expect high levels of default choice in small sets that do not include the strawberry jam, and low levels of default choice in both small sets that included the strawberry jam and the grand set. Thus, if a researcher randomly selected for analysis a small choice set without the strawberry jam, it would make it very hard to spot choice overload. Collecting data on all small choice sets and applying the Min bound would address this problem. 

The second issue is preference heterogeneity. Consider Example \ref{ex:min}: Here, all alternatives are equally likely to be attractive. Default choice probability is the same in all choice sets of the same size, and so a simple application of the Min bound will not increase power to detect choice overload. However, the RUM bound based on observations from choice sets of size two and three further constrains choice probabilities. (In a perfectly homogeneous population, the bounds coincide.)

A feature of the RUM (or indeed any MOD) bound is that it requires choice from $\mathcal{D}/X$ to be consistent with the relevant model. There are two possible issues with this. First, it could fail at the population level; in that case, $p_d^{RUM}(X)$ is not well-defined. Second, rationalizable population distributions will still, at least occasionally, generate non-rationalizable finite sample frequencies; in that case, one could define a feasible version of $p_d^{RUM}(X)$, and we will do so in Section \ref{sec:analysis}. 

Notice that this feature of the RUM bound, along with our focus only on default choice probabilities, means that many well-known behavioral phenomena will not show up as choice overload as per our definition -- either because they would lead to violations of RUM in the set $\mathcal{D}/X$, or because they would not be observable due to the coarsening of our data.  For example, consider a case in which the DM was first asked to choose between two hard to compare alternatives -- $x$ and $y$, say -- plus a default, and then asked to choose from the same set with an alternative $z$ that is dominated by $x$ but not $y$. The asymmetric dominance effect might push people to switch their choices from $y$ to $x$ in the larger choice set, but if this does not affect the choice of default, it will not show up as choice overload.

It remains to clarify how we can test stochastic rationalizability of data. For the case of standard stochastic choice data, this question has been resolved by \citet[][see \citet{Stoye19} for a short proof]{McFadden1991}. Here we adapt their approach to our data set. 

The basic insight is that choice probabilities can be rationalized if, and only if, they can be expressed as convex combination of data that would be produced by \textit{deterministic} utility maximizers. To make this precise, construct a matrix $\boldsymbol{A}$ s.t. each row of $\boldsymbol{A}$ corresponds to a given alternative \textit{within a particular choice set} (i.e., each alternative appears once for every choice set containing it). Each column corresponds to a deterministic choice pattern rationalizable by a different (strict) preference profile over $X$. For example, let $\mathcal{D}=\{\{a_{1},d\},\{a_{1},a_{2},d\},\{a_{1},a_{2},a_{3},d\}\}$ and let the first row of $\boldsymbol{A}$ indicate choice of $a_1$ from $\{a_1,d\}$, the second row choice of $d$ from $\{a_1,d\}$, and so on. One column of $\boldsymbol{A}$ (namely, the first one in \eqref{eq:A}) then represents the choices of someone who picked $a_1$ from all three choice sets, rationalizable by preferences ranking $a_1$ first.  Constructing the remaining columns from all other rationalizable choice patterns, one has:
\begin{equation} \label{eq:A}
\boldsymbol{A}=
\left[ 
\begin{array}{cccccccc}
1 & 1 & 1 & 1 & 0 & 0 & 0 & 0 \\ 
0 & 0 & 0 & 0 & 1 & 1 & 1 & 1 \\ 
1 & 1 & 0 & 0 & 0 & 0 & 0 & 0 \\ 
0 & 0 & 1 & 1 & 0 & 0 & 1 & 1 \\ 
0 & 0 & 0 & 0 & 1 & 1 & 0 & 0 \\ 
1 & 0 & 0 & 0 & 0 & 0 & 0 & 0 \\ 
0 & 0 & 1 & 0 & 0 & 0 & 1 & 0 \\ 
0 & 1 & 0 & 1 & 0 & 1 & 0 & 1 \\ 
0 & 0 & 0 & 0 & 1 & 0 & 0 & 0
\end{array}\right] 
\quad\begin{array}{l}
a_{1}\mid\left\{a_{1},d\right\}  \\ 
d\mid\left\{a_{1},d\right\}  \\
a_{1}\mid\left\{a_{1},a_{2},d\right\}  \\ 
a_{2}\mid\left\{a_{1},a_{2},d\right\}  \\ 
d\mid\left\{a_{1},a_{2},d\right\}  \\
a_{1}\mid\left\{a_{1},a_{2},a_{3},d\right\}  \\ 
a_{2}\mid\left\{ a_{1},a_{2},a_{3},d\right\}  \\ 
a_{3}\mid\left\{ a_{1},a_{2},a_{3},d\right\}  \\ 
d\mid\left\{ a_{1},a_{2},a_{3},d\right\}  
\end{array}.
\end{equation}

\citeauthor{McFadden1991}'s (\citeyear{McFadden1991}) core insight is the following. 
\begin{theorem}\label{MRTheorem}
Let the vector $\rho$ collect observed choice probabilities in order corresponding to the rows of $\boldsymbol{A}$. These probabilities are rationalizable by RUM if, and only if, there exists a vector $\nu \in \Delta^{H-1}$ (the $(H-1)$-dimensional unit simplex, where $H$ is the number of columns of $\boldsymbol{A}$) such that 
\begin{equation*}
\boldsymbol{A}\nu=\rho.
\end{equation*}%
\end{theorem}
We slightly adapt this approach because we only observe whether $d$ was chosen from any choice set; thus, the choice information encoded in $\boldsymbol{A}$ is too granular. To avoid this, one may in general premultiply $\boldsymbol{A}$ by a matrix $\boldsymbol{B}$ that merges choice of different non-default options in a given menu. Columns of $\boldsymbol{B}\boldsymbol{A}$ then represent different rationalizable deterministic ``default vs non-default'' choice patterns. 
For the example above, one would have
\begin{eqnarray*}
\boldsymbol{B}&=&\left[ 
\begin{array}{ccccccccc}
1 & 0 & 0 & 0 & 0 & 0 & 0 & 0 & 0 \\ 
0 & 1 & 0 & 0 & 0 & 0 & 0 & 0 & 0 \\ 
0 & 0 & 1 & 1 & 0 & 0 & 0 & 0 & 0 \\ 
0 & 0 & 0 & 0 & 1 & 0 & 0 & 0 & 0 \\ 
0 & 0 & 0 & 0 & 0 & 1 & 1 & 1 & 0 \\ 
0 & 0 & 0 & 0 & 0 & 0 & 0 & 0 & 1%
\end{array}%
\right] \\
\boldsymbol{BA}&=&\left[ 
\begin{array}{cccccccc}
1 & 1 & 1 & 1 & 0 & 0 & 0 & 0\\ 
0 & 0 & 0 & 0 & 1 & 1 & 1 & 1\\ 
1 & 1 & 1 & 1 & 0 & 0 & 1 & 1\\ 
0 & 0 & 0 & 0 & 1 & 1 & 0 & 0\\ 
1 & 1 & 1 & 1 & 0 & 1 & 1 & 1\\ 
0 & 0 & 0 & 0 & 1 & 0 & 0 & 0%
\end{array}%
\right]
\simeq\left[ 
\begin{array}{cccccccc}
1 &  0 & 0 & 0\\ 
0 &  1 & 1 & 1\\ 
1 &  0 & 0 & 1\\ 
0 &  1 & 1 & 0\\ 
1 &  0 & 1 & 1\\ 
0 &  1 & 0 & 0%
\end{array}%
\right]
\quad\begin{array}{l}
\text{not }d\mid\left\{a_{1},d\right\}  \\ 
d\mid\left\{a_{1},d\right\}  \\
\text{not }d\mid\left\{a_{1},a_{2},d\right\}  \\ 
d\mid\left\{a_{1},a_{2},d\right\}  \\
\text{not }d\mid\left\{a_{1},a_{2},a_{3},d\right\}  \\ 
d\mid\left\{ a_{1},a_{2},a_{3},d\right\}  
\end{array},
\end{eqnarray*}
where the last expression removes redundant columns of $\boldsymbol{BA}$; it will become clear that this simplification is without loss of generality. In specific applications, it may be possible (and easier) to directly enumerate columns of $\boldsymbol{BA}$, for example because the structure of a data set allows for a simple revealed preference characterization of admissible ``default/non-default'' choice patterns. This can be seen in the example above, where it is not difficult to directly verify that the simplified $\boldsymbol{BA}$ is the desired matrix. It is also the road we will take in our application.

In either case, a corollary to Theorem \ref{MRTheorem} characterizes $p_d^{RUM}(X)$. To state it formally, let the vector $\pi$ collect probabilities of active and passive choice in order corresponding to rows of $\boldsymbol{BA}$; in our example,
\begin{equation*}
\pi = \begin{pmatrix} 1-p_d(\{a_1,d\}) \\ p_d(\{a_1,d\}) \\ 1-p_d(\{a_1,a_2,d\}) \\ p_d(\{a_1,a_2,d\}) \\ 1-p_d(\{a_1,a_2,a_3,d\}) \\ p_d(\{a_1,a_2,a_3,d\}) \end{pmatrix}.
\end{equation*}
Then we can write:
\begin{corollary}\label{Corr1}
A probability vector $\pi$ as just defined is rationalizable by RUM if, and only if, there exists $\nu \in \Delta^{H-1}$ such that 
\begin{equation}
\boldsymbol{BA}\nu=\pi. \label{eq:McFR_problem}
\end{equation}
Further, let $\bm{a}$ be the row of $\bm{A}$ that corresponds to default choice from $X$, then
\begin{equation}
p_{d}^{RUM}(X) = \max_{\nu \geqq 0} \{\bm{a}\nu\} \text{    s.t.    }\boldsymbol{BA}\nu=\pi. \label{eq:p_RUM}
\end{equation}
\end{corollary}
For intuition, observe that in \eqref{eq:A}, the vector $\bm{a}$ is the last row of $\bm{A}$. Equivalently, it is the indicator of the unique choice type whose choice is always the default. The bound simply maximizes the probability of this type, subject to overall data being stochastically rationalizable.\footnote{Expression \eqref{eq:p_RUM} presupposes that $\pi$ is stochastically rationalizable. Since the set of rationalizable probability vectors is ``small'' (we will elaborate on this in Section \ref{sec:analysis}), empirical choice frequencies may fail this. In that case, a feasible version of the bound can be computed by substituting a constrained estimator of $\pi$. This will be illustrated later.} Note finally that, by inspection, the value of problem \eqref{eq:p_RUM} is not affected by removing redundant columns from $\boldsymbol{BA}$ (and correspondingly abridging $\bm{a}$). It is this last version of the problem that we solve in our application.   

\subsection{Generalizing the Testing Strategy}

We next build on these results to lay out a more general testing strategy that both clarifies the generality of our approach and provides a sense of rejecting \textit{in favor of a different model}, thereby identifying choice overload (as opposed to some general failure of the model) as the main concern. This generalization rests on two observations:
\begin{itemize}
\item[(1)]The ``benchmark model'' that we compare against in Theorem \ref{MRTheorem} need not be standard economic rationality. It can be any finite collection of ``admissible'' choice patterns across a given set of choice problems, whether weaker (as in Section \ref{sec:benchmark}), stronger (as in \cite{aguiar2023random}), or nonnested with utility maximization.
\item[(2)]Extending that idea, we can test nested models that allow for different degrees of choice overload. In our application, we fail to reject an adaptation of our benchmark model that allows for some measure of choice overload while still being restrictive. In particular, choice overload might not only happen in a grand set, but more generally as choice sets expand.
\end{itemize}
 We next elaborate both generalizations.
 
 \subsubsection{Changing the Benchmark Model}\label{sec:benchmark}
Our approach can be applied to other models of choice. As an example, consider the Rational Shortlist Method (RSM; \citet{manzini2007sequentially}). A simplified version of this model posits that a decision maker has two rationales, a linear order $P_{2}$ and an acyclic but possibly incomplete order $P_{1}$ on the grand set $X$.\footnote{\citeauthor{manzini2007sequentially}'s (\citeyear{manzini2007sequentially}) model considers any asymmetric $P_1$ and $P_2$. \citet{apesteguia2013choice} consider the restriction to acyclic relations. } These rankings are applied sequentially, so choice from $A \subset X$ is given by:
\[
\mathcal{C}(A,P_1,P_2) = \{x \in M(A,P_{1}) \mid x P_{2} y \quad \forall y \in M(A,P_{1})\}
\]
where 
\[
M(A, P_{1}) = \{z \in A \mid y P_{1} z \text{ for no } y \in A\}
\]
The idea is that DM uses $P_{1}$ to rule out some alternatives and then chooses the best remaining one according to $P_{2}$. We say say alternative $x$ blocks $y$ if $x P_1 y$. RSM embeds standard utility maximization because $P_1$ may be empty. As we assume $P_2$ is a linear order and $P_2$ is acylic, exactly one choice is produced from every choice set.

Analogously to RUM, we consider data being generated by a choice-set independent distribution over RSM models or a random RSM model. Default choice data $p_d : \mathcal{D} \rightarrow \Delta [0,1]$ from choice sets $\mathcal{D} \subset 2^X \setminus \emptyset$ is consistent with a \emph{random RSM} model if there exists a finite collection of RSM models $\mathcal{P}$ and a probability distribution $\rho$ over them such that for each $A \in \mathcal{D}$

$$p_d(A) = \sum_{(P_1,P_2) \in \mathcal{P}_1 \times \mathcal{P}_2 | d = \mathcal{C}(A,P_1,P_2)} \rho(P_1,P_2).$$

This model provides a different definition of choice overload than those given by monotonicity and RUM. That is, default choice data from smaller sets give a different bound on $p_d(X)$ (or any larger superset) under RSM than either $p_d^{min}(X)$ or $p_d^{RUM}(X)$. This bound $p_d^{RSM}(X)$ is weakly larger than $p_d^{RUM}(X)$ for any data as random RSM is a more permissive model than RUM, but can be less than $1$. Indeed, as we detail below, while we find strong choice overload effects with respect to RUM in our experimental data, we find no choice overload with respect to random RSM. We note that RSM does not require monotonicity with respect to the default choice (as is true in the example below).\footnote{For example, in the first row of Table \ref{table example 2} the default is chosen from $\{z\}$ and $\{x,y,z\}$ but not $\{y,z\}$.} Hence there is no clear ranking between $p^{RSM}_d(X)$ and $p_d^{min}(X)$ --- we can find examples where either is larger. We first look at a simple example.

\begin{example}\label{ex:rsm}

Let $X=\{x,y,z,d\}$ and consider the following default choice probability data.

\begin{table}[h!]
\centering
\begin{tabular}{ll}
\toprule
Choice set & $p_{d}$ \\
\midrule
$x, d$ &  $1/4$\\
$y, d$ &  $1/4$\\
$z, d$ &  $1/4$ \\
$x, y, d$ & $1/4$ \\
$x, z, d$ & $1/4$ \\
$z, y, d$ &  $1/4$ \\
\bottomrule
\end{tabular}
\caption{Hypothetical choice probabilities under random RSM.}\label{table example}
\end{table}

The RUM and Min bounds here equal $1/4$, yet $p_d^{RSM}(X)=3/4$. This bound is achieved by a population of four equiprobable RSM types tabulated in Table \ref{table example 2}. A somewhat tedious argument relegated to Lemma \ref{lemma:rsm_example} shows that the bound is also tight.

\begin{table}
\centering
\small
\begin{tabular}{lllllllll}
\toprule
$P_{2}$ & $P_{1}$ & $p_{d}(x)$ & $p_{d}(y)$ & $p_{d}(z)$ & $p_{d}(x,y)$ & $p_{d}(x,z)$ & $p_{d}(y,z)$ & $p_{d}(x,y,z)$ \\
\midrule
$x y d z$ & $x y \ z x$ & 0 & 0 & 1 & 0 & 1 & 0 & 1 \\
$x z d y$ & $z x \ y z$ & 0 & 1 & 0 & 0 & 0 & 1 & 1 \\
$z y d a$ & $y z \ x y$ & 1 & 0 & 0 & 1 & 0 & 0 & 1 \\
$x y z d$ & & 0 & 0 & 0 & 0 & 0 & 0 & 0 \\
\midrule
Probability & & $\frac{1}{4}$ & $\frac{1}{4}$ & $\frac{1}{4}$ & $\frac{1}{4}$ & $\frac{1}{4}$ & $\frac{1}{4}$ & $\frac{3}{4}$ \\
\bottomrule
\end{tabular}
\caption{A random RSM generating the choice probabilities in Table \ref{table example}.}\label{table example 2}
\end{table}
\end{example}

In Appendix \ref{app:rsm} we discuss how to test for choice overload against the benchmark of RSM in practice. As RSM admits a finite number of deterministic types, one can in principle use the approach in \ref{Corr1} with an $\bm{A}$-matrix with columns giving all deterministic choice patterns possible under RSM. As RSM is significantly more flexible that RUM, however, computing this $\bm{A}$-matrix is computationally intensive for our experimental data. In the appendix we provide techniques that only require computing part of the $\bm{A}$-matrix to make the test (conservatively) feasible. In contrast to RUM, random RSM is not rejected in our data.

\subsubsection{More General Notions of Choice Overload}

Up until now we have focused on a notion of choice overload which applies only to the grand set $X$. However our approach can also be used to consider choice overload that `kicks in' at any arbitrary set. Broadly speaking, our aim is to define choice overload at a given set $A$ as behavior such that the DM (i) chooses $d$ at set $A$ in violation of a particular model of choice and (ii) continues to choose $d$ at all supersets of $A$.

In our empirical application, we will not just test for evidence of choice overload in the grand set, but we will test a sequence of nested models. They are:
\begin{itemize}
\item[(i)] RUM.
\item[(ii)] A relaxation of RUM that allows for choosing $d$ from choice set $X$, regardless of behavior on smaller sets.
\item[(iii)] The same model as (ii), except that a choice may switch to $d$ for \emph{all} choice sets of cardinality $3$. (Types that choose $d$ from all sets of cardinality $3$ must also choose $d$ from $X$.)
\end{itemize}
All of these can be implemented by adding columns to $\bm{A}$ that corresponding to the additional choice types allows. The RUM bound is violated if (i) is rejected while model (ii) is not: Model (ii) captures choice overloaded behavior by only allowing violations of standard rationality through switching to $d$ when the choice set expands. 

Model (iii) expands on the standard notion of choice overload by allowing it to ``kick in'' at smaller choice sets. Moving from (i) to (iii) enlarges $\bm{A}$ but does not add conceptual difficulties. Since the resulting models are nested, one can then ask what is the least permissive model that is not rejected in the data. This will allow us to unpack what sort of choice overloaded behavior, if any, could have generated our data. In Section \ref{sec:analysis}, we show that model (iii) is the only one not rejected in our data and also argue that is still restrictive.\footnote{A non-nested relaxation of RUM would be to alow for choice overload as one moves from binary to ternary menus but not thereafter. This model is soundly rejected in our data.}

\subsection{Statistical Tests}

We next explain testing strategies that connect the above ideas to recent advances in econometrics.\footnote{This section can be skipped without loss of continuity.} To this purpose, we consider samples that were generated by randomly drawing individuals and then giving each individual a (i.i.d. randomly generated) selection of choice problems.\footnote{This mirrors our empirical design, which was partly chosen because, unlike stratified sampling or mean-reverting coins, it is easy to bootstrap.} In particular, data may contain choices from the same individuals in different choice sets, as is the case in our application.

We estimate choice probabilities $p_d(\cdot)$ by the analogous sample frequencies $\hat{p}_d(A)$. For all but the finite sample test that we present first, any estimator of $p_d(\cdot)$ whose asymptotic distribution is normal or approximated by the simple nonparametric bootstrap would suffice. Sample frequencies have both properties as long as they are not close to degenerate, where ``close to" is relative to sample size. We chose a relatively large sample size for our experiment precisely to ensure this, and this condition easily holds in our data.

\subsubsection{A Finite-Sample Test of the Min Bound}\label{sec:min_exact}

Testing the Min bound amounts to testing whether
\begin{eqnarray*}
&& p_d(X) \leq \min_{A \in \mathcal{D}\setminus X} p_d(A) \\
&\Longleftrightarrow & p_d(X) \leq p_d(A), \forall A \in \mathcal{D}\setminus X.
\end{eqnarray*}
The second expression clarifies that this is a joint test of potentially many hypotheses. Consider first testing any one of these, i.e. testing whether $p_d(X) \leq p_d(A)$ for a specific $A \subset X$. To this purpose, define the ``leave-$A$-out'' sample frequency $\hat{p}_{d,A}(X)$ by dropping observations from subjects who also saw choice problem $A$. As a result, $\hat{p}_d(A)$ and $\hat{p}_{d,A}(X)$ estimate binomial proportions in two mutually exclusive samples. We therefore apply \citeauthor{Fisher34}'s (\citeyear{Fisher34}) exact test for binomial proportions to $H_0:p_d(X)\leq p_d(A)$ for any given $A$.

Of course, we need to account for the fact that we conduct many such tests once ($78$ in our application) and cannot assume independence. Our first approach is Bonferroni adjustment, that is, all p-values are multiplied by $78$. An advantage of this approach is that it ensures finite-sample (as opposed to asymptotic) size control. However, its power is limited through three channels: The estimators $\hat{p}_{d,A}(\cdot)$ discard data; Fisher's exact test is in general conservative due to integer issues and a strong sense of conditional validity; Bonferroni adjustment is conservative. In practice, with the sample sizes that we generated for our empirical application, we expect only the last channel to have an appreciable effect. 

\subsubsection{An Asymptotic Test of the Min Bound}\label{sec:min_asy}

The finite sample test adjusts for the fact that, in principle, many tests are conducted simultaneously. A common concern with such adjustments is that, if the results of some of these tests appear obvious, one might needlessly lose power. Indeed, in our experimental data, the default probabilities in a number of choice sets are obviously much higher than in the grand choice set.\footnote{Figure \ref{fig:histogram} in Supplemental Appendix \ref{app:choice} provides a visualization of this.} Can we restrict attention to only those inequality conditions that might reasonably bind?

This question has received considerable attention in the econometric literature on moment inequalities. We implement a method that can be seen as special case of \citet{AS10} and also of \citet{CLR13}, both of whom establish its validity under rather general conditions. 

The method can use many test statistics; for concreteness, set $$t=\hat{p}_d(X)-\min_{A \in \mathcal{D}\setminus X} \hat{p}_d(A).$$
The test will reject if $t$ is too large. The catch is that the distribution of $t$, and therefore the appropriate critical value, depends on the nuisance parameter $(p_d(X)-p_d(A))_{A \in \mathcal{D}\setminus X}$. This parameter cannot be pre-estimated with sufficient accuracy\footnote{Technically, it enters the asymptotic distribution scaled by $\sqrt{n}$. See \citet[][ section 4]{CS17} for a survey.} and so we must conservatively approximate it. This is done in three steps:
\begin{enumerate}
\item Use the simple nonparametric bootstrap to approximate the distribution of $$(\hat{p}_d(A)-p_d(A))_{A \in \mathcal{D}}$$
by the (bootstrap) distribution of
$$(\tilde{p}_d(A))_{A \in \mathcal{D}} \equiv (\hat{p}_d^*(A)-\hat{p}_d(A))_{A \in \mathcal{D}},$$
where $\hat{p}_d^*(\cdot)$ denotes the bootstrap analog of $\hat{p}_d(\cdot)$. This bootstrap will be clustered by individual, i.e. we (i.i.d. uniformly with replacement) resample individuals and use all responses from a given resampled individual; this ensures that correlation patterns in $\hat{p}_d(\cdot)$ due to eliciting several responses per individual are captured.
\item Use a pre-test with size converging to $0$, e.g. $\alpha_n=\alpha/\log(n)$, where $\alpha$ is the test's nominal size. Discard from consideration any sets $A$ s.t. the null hypothesis $H_0: p_d(X) \geq p_d(A)$ is rejected at significance level $\alpha_n$. Let $\mathcal{D}^*$ denote the set of choice problems that are retained in this pre-test.
\item The critical value of our test is the appropriate quantile of the recentered bootstrap test statistic
$$t^* \equiv \tilde{p}(X)-\min_{A \in \mathcal{D}^*\setminus X} \tilde{p}(A).$$
\end{enumerate}
This procedure reflects two important ideas from the moment inequalities literature. First, the bootstrap population of data must be on the null hypothesis, which necessitates a recentering. In our case, the least favorable and therefore relevant instance of the null hypothesis is that all relevant probabilities are equal. Since the test statistic is location invariant, for concise notation and implementation we recenter them to $0$. This is reflected in the definition of $\tilde{p}_d(\cdot)$. Second, the test may be extremely conservative if we accordingly recenter all $78$ estimators. Therefore, we pre-screen choice items whose default probability is likely to much exceed $p_d(X)$. In particular, because the size of the pre-test goes to $0$, we will asymptotically select all binding constraints.\footnote{Our depiction of the method is simplified by picking a specific test statistic and also filling in specific values for several tuning parameters. Also, rather than letting the size of a pre-test vanish, one could use Bonferroni correction to ``spend'' some ``coverage budget'' on the pre-test \citep{AndrewsJia,RSW}. This would make no difference in our empirical application because its p-values are far from conventional thresholds.}

\subsubsection{An Asymptotic Test of RUM and its Generalizations}\label{sec:RUM_test}

 
Statistical testing of Random Utility Models is due to \cite{KS18}, with important computational improvement by \citet{RUM_compute}. It has seen application to observational data \citep{DKQS16} as well as lab experiments \citep{aguiar2023random}. Recall the approach only requires that the population is modeled as mixing a finite number of ``admissible" types encoded in the columns of $\bm{A}$, not that those types coincide with standard economic rationality. Hence, we can use the machinery to test nonstandard and, in particular, nested models.

\paragraph{The Hypothesis Test} 
A main insight in \citet{KS18} is that the null hypothesis
$$H_0: \boldsymbol{BA} \nu = \pi, \exists \nu \in \Delta^{H-1}$$
can equivalently be written as
\begin{eqnarray*}
&H_0: \min_{\nu \geq 0} \{(\pi-\boldsymbol{BA}\nu)'\Omega(\pi-\boldsymbol{BA}\nu)\}=0,
\end{eqnarray*}
where $\Omega$ is an arbitrary positive definite (and in practice diagonal) weighting matrix. That is, the residuals from projecting $\pi$ onto the cone $\mathcal{C}$ of rationalizable probabilities must equal $0$.  This suggests the scaled norm of the corresponding sample residuals as test statistic. Noting the similarity to specification tests in multiple equation models \citep{Sargan58,Hansen82}, call this statistic
\begin{equation*}
J_n \equiv n \min_{\nu\geq 0} \{(\hat{\pi}-\boldsymbol{BA}\nu)'\Omega(\hat{\pi}-\boldsymbol{BA}\nu)\},
\end{equation*}
where $\hat{\pi}$ is the sample analog of $\pi$ and $n$ is sample size.

Despite the superficial similarity to well-established methods, the asymptotic distribution of $J_n$ is hard to estimate because it depends discontinuously on where on $\mathcal{C}$ the true $\pi$ is. However, one main contribution of \citet{KS18} is precisely to overcome this problem. Following them, we approximate the distribution of $J_n$ by the one of a modified bootstrap analog
\begin{eqnarray}
J^*_n &\equiv &  \min_{\nu\geq \bm{1}\cdot\tau_n/H} \{(\hat{\pi}_{\tau_n}^*-\boldsymbol{BA}\nu)'\Omega(\hat{\pi}_{\tau_n}^*-\boldsymbol{BA}\nu)\} \label{eq:J*N} \\
\hat{\pi}_{\tau_n}^* &\equiv & \hat{\pi}^* + \hat{\eta}_{\tau_n} - \hat{\pi} \notag \\
\hat{\eta}_{\tau_n} &\equiv & \arg\min_{\nu\geq \bm{1}\cdot\tau_n/H} \{(\hat{\pi}-\boldsymbol{BA}\nu)'\Omega(\hat{\pi}-\boldsymbol{BA}\nu)\}, \label{eq:eta}
\end{eqnarray}
where $\tau_n$ is a tuning parameter that we set in accordance with the literature,\footnote{Specifically, $\tau_n=\sqrt{\log(\underline{n})/\underline{n}}$, where $\underline{n}= \tfrac{2qn}{k(k+1)}$ is expected sample cell size for any but the universal choice problem; here, $k$ is the number of nondefault items in $X$ and $q$ is the number of ``small" choice problems faced by each subject. Recall also that $H$ is the length of $\nu$, thus division by $H$ ensures that the above constraint is scaled by $\tau_n$ and not by the testing problem's complexity.} $\bm{1}$ is a vector of $1$'s, and $\hat{\pi}^*$ is a simple nonparametric (clustered, as explained earlier) bootstrap analog of $\hat{\pi}$. 

We go beyond a completely straightforward implementation of \citeauthor{KS18}'s (\citeyear{KS18}) test because we do not weight questions equally. This possibility is anticipated by \citet{KS18}, who only assume a diagonal weighting matrix $\Omega$, but has not, to our knowledge, been implemented before. We use it because, in our data, choice probabilities pertaining to the universal choice set $X$ will be estimated from a much larger sample cell than others. 
We take this into account by weighting estimated probabilities for different questions by the expected sample cell size for that question; in practice, that means to weight all small questions equally and to put weight $w \approx 9$ on choice frequencies corresponding to the grand set.\footnote{More precisely, $w=\tfrac{k(k+1)}{2q}$, with $(k,q)$ as in the previous footnote. We do not weight questions by \textit{realized} sample cell sizes, and we also do not estimate cell-specific variances by the binomial variance formula, in order to avoid data dependent weighting. In our empirical application, these modifications would have minimal effect.}

In general, this test can be expensive to compute. The experimental design that we settled on, partly to ensure reasonable sample cell sizes, is small enough so that this concern does not arise. However, in preparation, we also implemented an adaptation of the computational improvements in \citet{RUM_compute}. These details are laid out in Supplementary Appendix \ref{app:compute}.

\section{An Application to Experimental Data}\label{sec:Experiment}

\subsection{Experimental Design}
Our testing strategy requires observing choices from a grand set of alternatives and a number of subsets. Based on the findings of \citet{chernev2015choice}, we want the choice problems to be non-trivial to increase the probability of finding choice overload. To this end, we ask subjects to make choices between amounts of experimental points expressed as sums, where each non-default option is expressed as sum of four numbers between $0$ and $10$ written in text. The value of choosing an option in experimental points is the value of the sum, and one experimental point is worth $50$ cents. There are $12$ non-default options in the grand set $X$. We generate these by first drawing the value of an alternative from an exponential distribution truncated at 10 points with $\lambda=0.25$, then randomly selecting individual terms of the sum so that neither the first nor the maximal summand was correlated with the total value of the option. (All of this follows \citet{caplin2011search}.)

\begin{figure}
\centering

\subfigure[]{\includegraphics[width=0.49\textwidth]{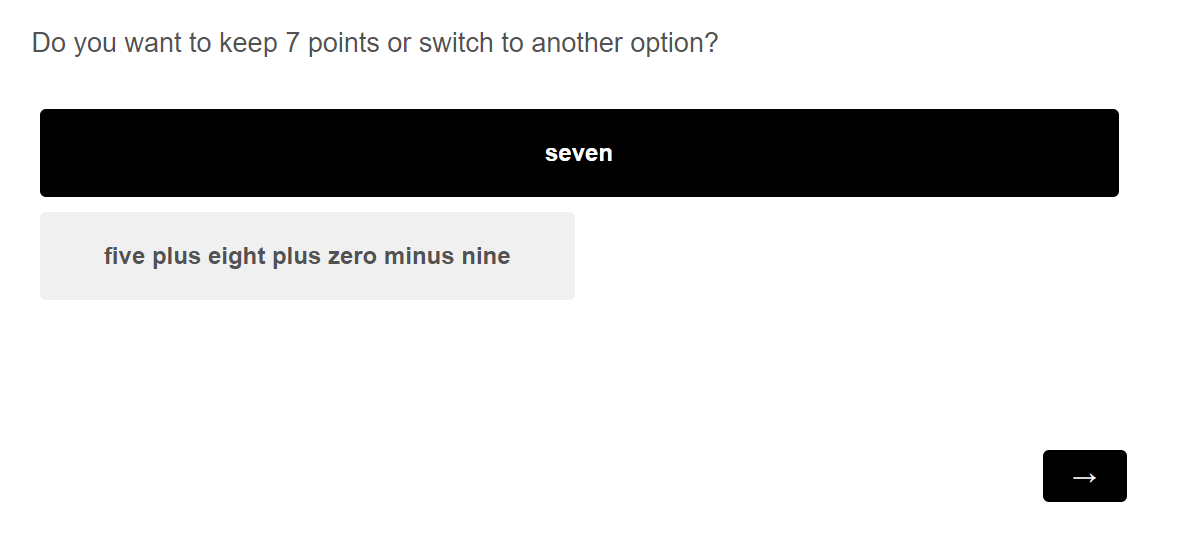}}
\subfigure[]{\includegraphics[width=0.49\textwidth]{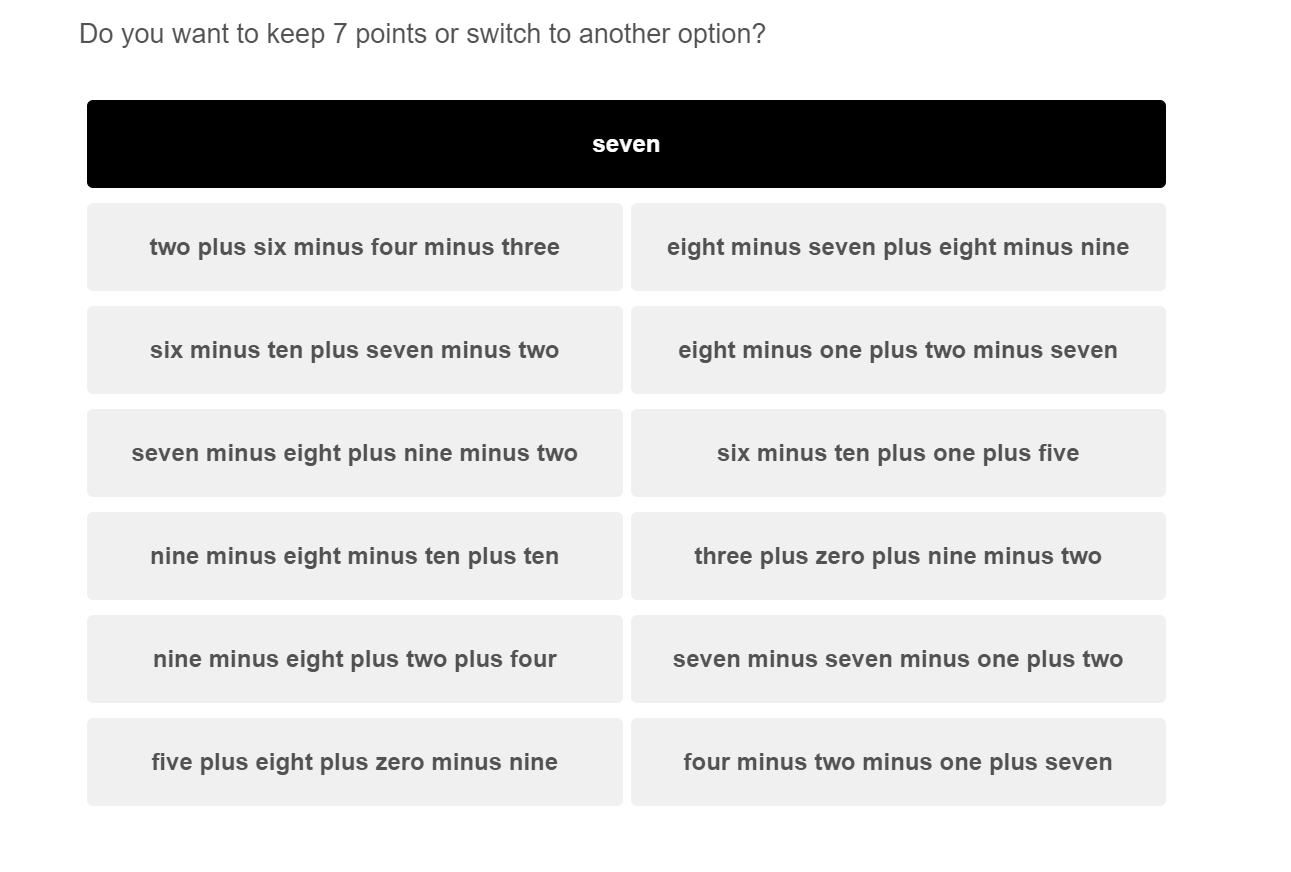}}

\captionof{figure}{Choice problems: (a) comparing one lottery to the default and (b) the grand set to the default.}
\label{fig:1}

\end{figure}

Each choice set also contained a default option providing $7$ experimental points, expressed as a single number. This option furthermore appeared at the top of the screen and was pre-selected, so it was an obvious default choice. Of the non-default options, $9$ yielded a number of points strictly lower than the default, $1$ yielded the same number of points, and $2$ yielded strictly more points. Figure \ref{fig:1}(a) shows an example choice screen with the default and one other option. Figure \ref{fig:1}(b) shows an example choice screen for the grand set.

The collection of choice sets consisted of the grand set and all subsets containing 1 or 2 alternatives along with the default, for a total of 78 smaller choice sets. Based on the prior literature, we initially believed that 3-item sets are unlikely to trigger choice overload, while 13-item sets are likely to do so if such an effect is present to begin with. Each subject was presented with 9 randomly selected small sets and the grand set, with the order of choice questions and the order of non-default options in each choice question randomized. One question was randomly selected for payment. Subjects in addition received a \$1 participation fee.  A complete list of choice alternatives and choice sets is in Supplementary Appendix \ref{app:choice}.\footnote{The experiment was approved by the Institutional Review Boards of Columbia and Cornell Universities.}

We note that this is a situation in which we would anticipate item heterogeneity to outweigh preference heterogeneity -- assuming people like more money to less, there should not be any preference heterogeneity. This means that, as per our discussion in Section \ref{sec:analysis}, we would expect the Min bound to perform well relative to the RUM bound.

\textbf{Subject Recruitment.} The experiment was run on Amazon's Mechanical Turk (MT) platform. This platform was chosen to easily collect data from a large number of subjects, each answering a small number of questions. ``Requesters'' post Human Intelligence Tasks (``HITs") -- usually simple jobs that pay small sums for each completed task. Workers on MT view descriptions of the HITs, decide which to accept, and complete those HITs over the internet. In order to improve subject attentiveness and reduce the probability of responses from bots, we prescreened subjects using the platform CloudResearch. This has been shown to be effective in increasing data quality \citep{chandler2019online, litman2020conducting}.

We recruited 2000 subjects in April and May of 2022. In addition to CloudResearch's screening, we restricted recruitment to MTurk workers who had completed over 1000 HITs and had an approval rating of over 97\%. 1833 subjects passed a comprehension quiz\footnote{Screenshots of the quiz and instructions can be found in Supplemental Appendix \ref{app:quiz}.} and completed the experiment, but one of these subjects had a browser-related error which made their data unusable, leaving us with 1832 subjects' data. In addition to 1832 choices from the grand set, we have at least 185 observations of choices from each of the small sets, with variation in sample size due to the random selection of choice questions. Supplementary Appendix \ref{app:choice} lists default choice frequencies for each choice set.

\subsection{Analysis} \label{sec:analysis}
We test for choice overload in roughly increasing order of presumed test power. To showcase the general applicability of our methods we ignore some features of our data at hand. Most notably, we do not make use of the fact that utility is essentially observable in our experiment, assuming that more money is better than less. While this would sharpen the tests we could perform, such information is generally not available, so we demonstrate how our tests work without making use of this information.

\paragraph{Existing Tests for Choice Overload}First, classic tests from the literature would be unlikely to find evidence of choice overload. The frequency of default choice in the grand set was $22\%$; the analogous frequency across all subsets was $71\%$. A simple comparison of means would, therefore, not reveal choice overload, and a random selection of a single smaller set would also be unlikely to do so -- only 15 of 78 (19\%) small sets have default choice significantly below the grand set's.

\paragraph{The Min Bound} Next, we use the finite sample test to ask whether any single default frequency is significantly below the one in the grand choice set, taking into account sampling uncertainty and multiple testing. Strikingly, the answer is yes: After Bonferroni adjustment, the p-value against the null hypothesis that the grand choice set default probability is lowest is $.00003$, and $5$ choice items are significantly lower at the $5\%$-level. Similarly, the asymptotic test of the Min bound yields a p-value that we could not distinguish from $0$ in $B=10000$ Monte Carlo simulations. Hence, we find strong evidence of choice overload using this approach.


\paragraph{The RUM Bound}
We next test consistency with RUM. At a Monte Carlo replication size of $B=10000$, the p-value equals $0$ as well. This strong rejection comes with a caveat: The p-value against all but the grand choice set equals $.005$. Therefore, ``the data are consistent with RUM except that default choice from $X$ is too frequent" is not an appropriate description of our findings. However, this leaves open the possibility that, contrary to our own prior, choice overload had an effect in some of the smaller choice sets.

To test this, we next consider models (ii) and (iii) from Section \ref{sec:population} by running the test from Section \ref{sec:RUM_test} with appropriately modified $\bm{A}$-matrices. The p-value using all data but applying extension (ii), i.e. allowing for choice types that are rationalizable but choose $d$ from $X$, is also $0.005$.\footnote{This may appear obvious from the preceding paragraph's results because economically, model (i) restricted to choice sets excluding $X$ is equivalent to model (ii). However, $p$-values need not be numerically the same due to subtleties of how the testing problem gets regularized. But one would expect them to be very similar (or else doubt the approximations involved), and their unrounded values in our data and using identical bootstrap draws are indeed $.0047$ vs $.0046$.} In contrast, the more general model (iii), i.e., allowing for subjects to switch to $d$ at $X$ or at all choice sets of size $3$ and up, is not rejected ($p=.65$). Given the details of our testing procedure, this also implies that the null hypothesis corresponding to the more restrictive model (ii) would be rejected while imposing the less restrictive model (iii).\footnote{This null hypothesis is linear (it can be written as $\bm{e}'\nu=0$, where the vector $\bm{e}$ is an indicator vector of columns of $\bm{A}$ that would reveal choice overload) and therefore can be tested using results from \citet{DKQS16}. Close inspection reveals that that test will numerically coincide with a direct test of the more restrictive model.} Subjects' behavior, therefore, appears to reveal choice overload already at sets of size $3$ (as well as at the grand set).\footnote{The Min bound test also indicates a failure of monotonicity between size 2 and 3 sets. While the most conservative of our tests, based on Fisher exact tests and Bonferroni corrections, just fails to reject the null hypothesis of monotonicity at the 5\% level, the asymptotic test of section 3.2.2 provides a clear rejection.} As prior experiments in the literature generally looked for choice overload at larger set sizes, it is economically interesting that subjects may be choice overloaded when facing such small sets.\footnote{\cite{tversky1992} report an increase in default choice when switching from size $2$ to size $3$ choice sets, but ascribe this to the disjunction effect rather than choice overload.}

We close with four additional observations. First, our data speaks so loudly that, in hindsight, our more sensitive tests were not needed. As an informal illustration of the new tests' potential, we replicated the entire analysis on the first $50\%$ of subjects. As expected, all p-values crept up. Of note, model (ii) above, previously rejected with $p=.005$, was no longer rejected ($p=.250$).
Furthermore, while the p-value associated with the asymptotic Min bound test (see Section \ref{sec:min_asy}) remains effectively $0$, the one associated with the exact test (see Section \ref{sec:min_exact}) is above $1\%$. The use case for the asymptotic tests would therefore have been more striking if we had collected only half the data. 

Second, to further quantify the sense in which these models fit the data, we compute the largest sample proportion of subjects such that individually rational behavior by these subjects would be compatible with empirical choice frequencies. This proportion is defined by the linear program
$$\max \bm{1}'\nu\text{     s.t.  }\bm{A}\nu \leq \hat{\pi}$$
and equals $1$ if, and only if, sample frequencies are rationalizable in terms of the behavior encoded in $\bm{A}$. This fraction is $.866$ for the RUM, increases to $.877$ for model (ii), and equals $.915$ for model (iii).

Third, one may worry that model (iii) is just not very restrictive, while maybe models (i) and (ii) are. This cannot be literally true because our test statistics are positive in all three tests, hence empirical choice frequencies do not conform to any model. But it could be a practical concern, notably if all possible data sets are close to the model. We investigated this through analyses inspired by \citet{Bronars}, \citet{Selten}, and \citet{BC11}. Essentially, these approaches propose comparing the test statistic observed in the data to that that observed in randomly generated pseudo data that matches the true data in some regard (for example, the choice sets from which choices are observed). Specifically, for models (i)-(iii) we calculated the expected mean square error (MSE) when matching data that were generated from the uniform distribution on $[0,1]^{79}$. The resulting values are 0.25 for model (i), 0.23 for (ii) and 0.21 for (iii). These numbers are all close to each other and far above the MSE when these models are applied to the data (which range from 0.03 to 0.01), showing that even our most permissive model places significant restrictions on the data.

Finally, the data can be used to illustrate the difference between the Min and RUM bounds. While empirical choice frequencies do not imply a well-defined $p_d^{RUM}(X)$, one can easily compute the vector $\hat{\eta}$ of choice probabilities that are closest to the empirical ones while being rationalizable; in fact, this computation is a by-product of the statistical tests. This allows us to compute a feasible analog of $p_d^{RUM}(X)$. Its value in our data is $9.8\%$. To compare apples to apples, we report that the same $\hat{\eta}$ implies a Min bound of $p_d^{min}(X)=11.4\%$. Once again, using the full implications of RUM is potentially much more informative than just testing monotonicity.

\section{Conclusion}

In this paper, we argued that existing tests for choice overload are not very sensitive, potentially explaining the ambiguous picture that emerges from the current literature. We proposed that, by collecting more data and fully using restrictions from economic theory, one can design better tests. We find choice overload in a novel data set where standard approaches (with the data collected for them) would have been unlikely to do so. Indeed, our richer data speak so loudly that with hindsight, our econometric innovations would not have been necessary to detect choice overload. We believe that the innovations are of interest nonetheless, and we also note that such things are bound to occur if one genuinely designs the empirical strategy before collecting data (as we did). 

We hope that our work will have three consequences. First, by providing a higher powered test of choice overload, it should clear up the question of whether this is indeed a real phenomenon. Second, given that (we suspect) it will show choice overload to be more prevalent than previously thought, we hope it will spur further theoretical and policy work designed to understand its causes and mitigate its effects. Finally, by providing a better tool for measuring when choice overload does occur, we hope it will facilitate the above work by providing a better empirical basis on which to theorize.

\bibliography{overload}

\newpage

\appendix
\newpage
\setcounter{page}{1}

\section{Supplementary Appendix: Experimental Details}

\subsection{Instructions and Quiz}\label{app:quiz}

\begin{figure}
\centering
\includegraphics[scale=0.8]{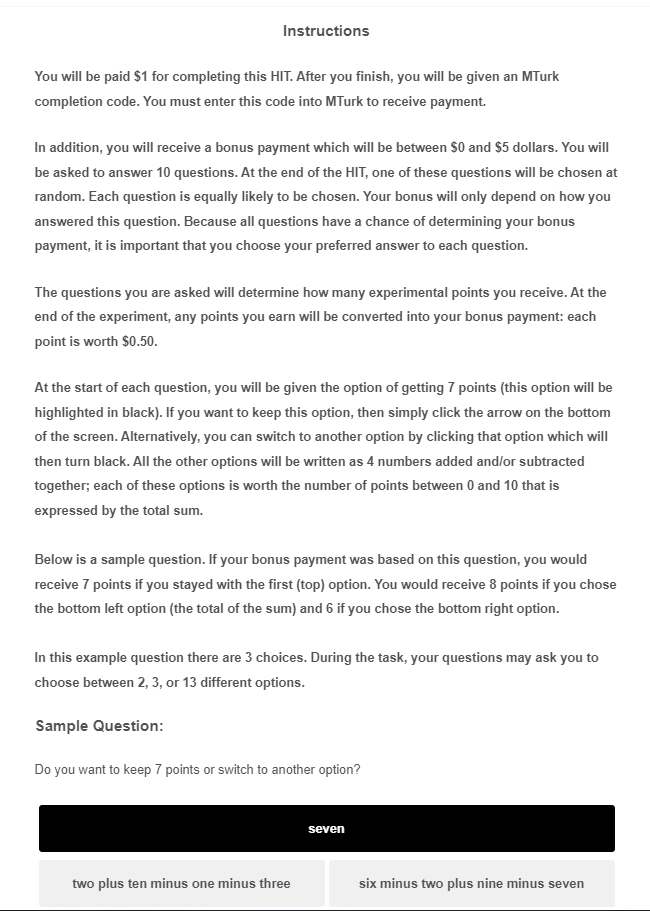} 
\caption{Instructions.}
\label{fig:instructions}
\end{figure}

Subjects were first shown an instructions screen before proceeding to the quiz; see Figure \ref{fig:instructions}. Subjects were given two attempts.

\begin{figure}
\centering
\includegraphics[scale=1.2]{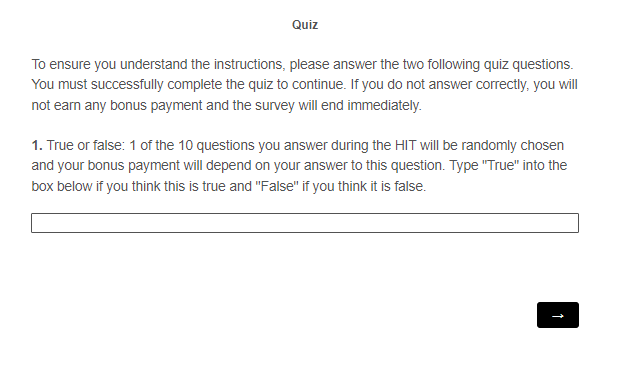} 
\caption{Quiz screen 1.}
\label{fig:quiz1}
\end{figure}

\begin{figure}
\centering
\includegraphics[scale=1.2]{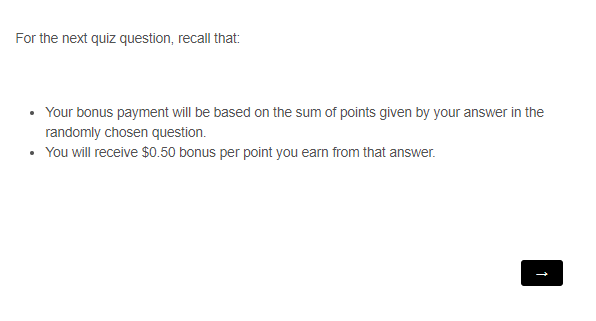} 
\caption{Quiz screen 2.}
\label{fig:quiz2}
\end{figure}

\begin{figure}
\centering
\includegraphics[scale=0.9]{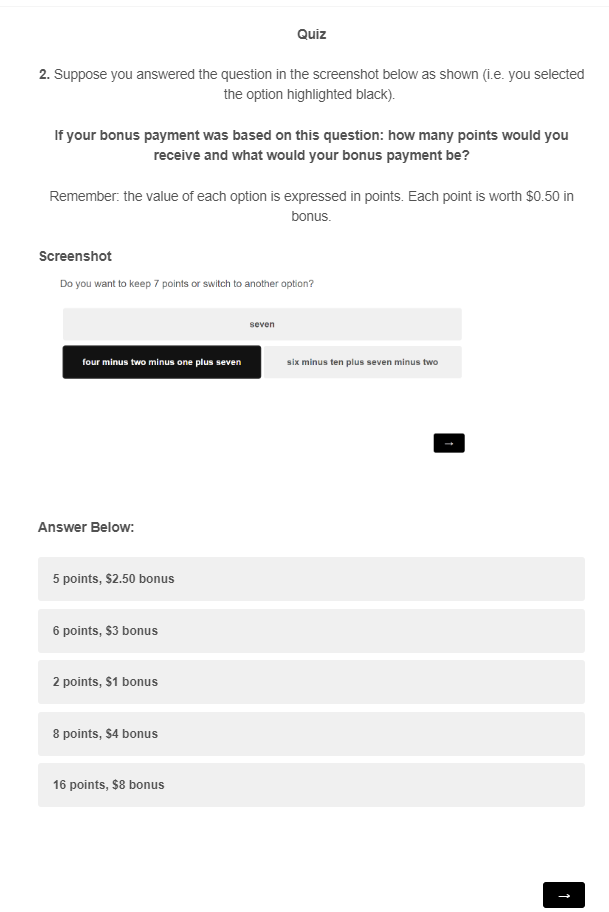} 
\caption{Quiz screen 3.}
\label{fig:quiz3}
\end{figure}

\subsection{Choice Alternatives with Summary Data}\label{app:choice}

\begin{figure}
\centering
\includegraphics[scale=0.8]{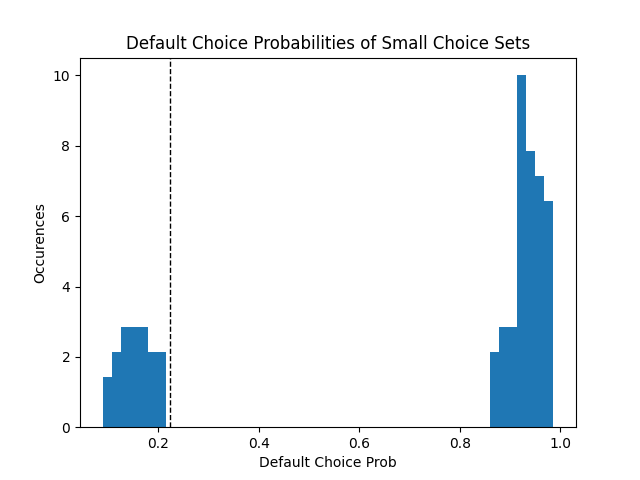}
\caption{Histogram showing distribution of choice sets by probability that default option was chosen. Dotted line shows the fraction of default choice in the grand set $X$.}
\label{fig:histogram}
\end{figure}

Table \ref{options_table} gives a full list of the choice objects with their values in experimental points. Table \ref{choice_sets_table} shows how often each choice set was shown to subjects and how often the default was chosen from it. Choice sets are given by the alternatives they contain: e.g. $[0,1,2]$ represents the choice set with the default (0) and options with IDs 1 and 2 from Table \ref{options_table}. Figure \ref{fig:histogram} shows a histogram of the default choice probabilities of all choice sets.

\begin{table}
\centering
\begin{tabular}{ c c c }
\textbf{ID} &  \textbf{Option} & \textbf{Value} \\
0 & seven & 7 \\
1 & eight minus seven plus eight minus nine & 0 \\
2 & eight minus one plus two minus seven & 2 \\
3 & seven minus seven minus one plus two & 1 \\
4 & six minus ten plus one plus five & 2 \\
5 & seven minus eight plus nine minus two & 6 \\
6 & five plus eight plus zero minus nine & 4 \\
7 & nine minus eight plus two plus four & 7 \\
8 & nine minus eight minus ten plus ten & 1 \\
9 & four minus two minus one plus seven & 8 \\
10 & two plus six minus four minus three & 1 \\
11 & three plus zero plus nine minus two & 10 \\
12 & six minus ten plus seven minus two & 1 
\end{tabular}
\caption{List of Options with Values}
\label{options_table}
\end{table}

\begin{table}
\centering
\begin{tabular} {ccc|ccc}
\textbf{Choice Set} &  \textbf{\# Choices} & \textbf{\# Default} & \textbf{Choice Set} &  \textbf{\# Choices} & \textbf{\# Default}   \\
$[0, 1]$ & 204 & 199 & $[0, 3, 10]$ & 199 & 190 \\
$[0, 2]$ & 193 & 188 & $[0, 3, 11]$ & 200 & 38 \\
$[0, 3]$ & 218 & 210& $[0, 3, 12]$ & 231 & 219\\
$[0, 4]$ & 232 & 225 & $[0, 4, 5]$ & 219 & 190\\
$[0, 5]$ & 227 & 214 & $[0, 4, 6]$ & 215 & 200\\
$[0, 6]$ & 230 & 226& $[0, 4, 7]$ & 213 & 191\\
$[0, 7]$ & 221 & 212& $[0, 4, 8]$ & 216 & 187\\
$[0, 8]$ & 229 & 212& $[0, 4, 9]$ & 193 & 35\\
$[0, 9]$ & 201 & 18& $[0, 4, 10]$ & 219 & 204\\
$[0, 10]$ & 199 & 190& $[0, 4, 11]$ & 210 & 28\\
$[0, 11]$ & 194 & 20& $[0, 4, 12]$ & 197 & 186\\
$[0, 12]$ & 195 & 184& $[0, 5, 6]$ & 224 & 200\\
$[0, 1, 2]$ & 209 & 203& $[0, 5, 7]$ & 209 & 183\\
$[0, 1, 3]$ & 225 & 217& $[0, 5, 8]$ & 210 & 186\\
$[0, 1, 4]$ & 185 & 175& $[0, 5, 9]$ & 224 & 45\\
$[0, 1, 5]$ & 204 & 190& $[0, 5, 10]$ & 199 & 189\\
$[0, 1, 6]$ & 208 & 199& $[0, 5, 11]$ & 214 & 36\\
$[0, 1, 7]$ & 203 & 188& $[0, 5, 12]$ & 213 & 199\\
$[0, 1, 8]$ & 225 & 203& $[0, 6, 7]$ & 205 & 189\\
$[0, 1, 9]$ & 211 & 24& $[0, 6, 8]$ & 200 & 178\\
$[0, 1, 10]$ & 218 & 215& $[0, 6, 9]$ & 218 & 44\\
$[0, 1, 11]$ & 223 & 30& $[0, 6, 10]$ & 209 & 204\\
$[0, 1, 12]$ & 235 & 219& $[0, 6, 11]$ & 223 & 31\\
$[0, 2, 3]$ & 229 & 219& $[0, 6, 12]$ & 221 & 210\\
$[0, 2, 4]$ & 213 & 202& $[0, 7, 8]$ & 223 & 202\\
$[0, 2, 5]$ & 218 & 202& $[0, 7, 9]$ & 206 & 36\\
$[0, 2, 6]$ & 215 & 208& $[0, 7, 10]$ & 199 & 182\\
$[0, 2, 7]$ & 250 & 231& $[0, 7, 11]$ & 205 & 30\\
$[0, 2, 8]$ & 193 & 178& $[0, 7, 12]$ & 221 & 205\\
$[0, 2, 9]$ & 207 & 36& $[0, 8, 9]$ & 226 & 32\\
$[0, 2, 10]$ & 192 & 185& $[0, 8, 10]$ & 205 & 182\\
$[0, 2, 11]$ & 194 & 24& $[0, 8, 11]$ & 222 & 33\\
$[0, 2, 12]$ & 192 & 182& $[0, 8, 12]$ & 221 & 204\\
$[0, 3, 4]$ & 191 & 182& $[0, 9, 10]$ & 192 & 31\\
$[0, 3, 5]$ & 218 & 203& $[0, 9, 11]$ & 202 & 23\\
$[0, 3, 6]$ & 198 & 194& $[0, 9, 12]$ & 223 & 42\\
$[0, 3, 7]$ & 205 & 192& $[0, 10, 11]$ & 219 & 33\\
$[0, 3, 8]$ & 215 & 194& $[0, 10, 12]$ & 193 & 187\\
$[0, 3, 9]$ & 207 & 44& $[0, 11, 12]$ & 224 & 36\\
&&& $[0,\ldots,12]$ & 1832 & 409
\end{tabular}
\caption{Choice Set and Default Choice Frequencies}
\label{choice_sets_table}
\end{table}

\section{Supplementary Appendix: \newline Implementing Recent Computational Innovations}\label{app:compute}
In the process of writing this paper, we implemented the computational procedure in \citet{RUM_compute}. To our knowledge, this is the first such implementation beyond their own illustrative example. The implementation and some not entirely obvious modifications are described next. This material is not in the main text because we did not end up using the implementation in the empirical work.

\citeauthor{RUM_compute}'s (\citeyear{RUM_compute}) procedure is motivated by the fact that computation of the matrix $\boldsymbol{A}$ and also computation \eqref{eq:J*N} is hard, and yet the latter needs to be repeated many times. It exploits that, because $\boldsymbol{BA}$ has many more columns than rows, there always exists a sparse (in the loose sense of having relatively few nonzero entries) $\arg\max$ to problem \eqref{eq:J*N}. We will avoid solving \eqref{eq:J*N} as stated, or ever computing $\boldsymbol{BA}$ (though the latter is feasible here), by guessing the nonzero entries. Formally, this goes as follows. (We drop $N$ for ease of notation.):
\begin{enumerate}
\item Initialize the matrix $\tilde{\boldsymbol{BA}}$ by constructing relatively few columns of $\boldsymbol{BA}$.
\item Compute $$\tilde{J} \equiv \min_{\nu\geq 0} \{(\hat{\pi}-\tilde{\boldsymbol{BA}}\nu)'\Omega(\hat{\pi}-\tilde{\boldsymbol{BA}}\nu)\}.$$ Let  $\tilde{\eta} \equiv \tilde{\boldsymbol{BA}}\tilde{\nu}$, where $\tilde{\nu}$ solves this problem. (While $\tilde{\nu}$ may not be unique,  $\tilde{\eta}$ is.) 
\item Maximize $(\hat{\pi}-\tilde{\eta})'\Omega(a-\tilde{\eta})$ subject to the constraint that $a$ is a column of $\boldsymbol{BA}$. This is called the ``pricing problem." Its constraint must be expressed in an application specific, computable way, and we do so below.
\item If the value of the problem just solved is positive, append column $a$ to $\tilde{\boldsymbol{BA}}$. Repeat until the value of the problem is nonpositive or another convergence criterion is met.
\end{enumerate}
The basic idea is that, as long as the deficient matrix $\tilde{\boldsymbol{BA}}$ does not contain all columns that receive positive weight in one solution to the original problem, the value of the simplified problem can be improved by appending such a column. But a column improves this value iff the supporting hyperplane separating the current feasible set from $\hat{\pi}$ does not separate the new column from $\hat{\pi}$. The program in step 3 simply checks this. (We solve it but in principle, it suffices to sign its value.) If the solution is sparse, it will be found while only generating a fraction of all possible columns of $\boldsymbol{BA}$.

Our implementation is again not completely off the shelf. Modifications are as follows:
\begin{itemize}
\item [(i)] We take account of the weighting matrix $\Omega$ not being the identity matrix. This is already reflected in expressions above.
\item[(ii)] The requirement that the vector $a$ be a possible column of $\boldsymbol{BA}$ can be expressed by writing the pricing problem as follows. To enforce that $a$ is binary and any two entries corresponding to the same choice problem sum to $1$, parameterize it in terms of a vector $\rho$ that only collects indicators of active choice. Then $a=d+D\rho$, where
\begin{eqnarray*}
d&=&\left(\begin{array}{c}
0 \\
1  \\
0  \\
\vdots \\
1
\end{array}\right), ~~~
D=\left(\begin{array}{rrrr}
1 & 0 & \dots & 0 \\
-1 & 0 & \dots & 0 \\
0 & 1 & \dots & 0 \\
\vdots & \vdots & \ddots & \vdots \\
0 & 0 & \dots & -1
\end{array}\right).
\end{eqnarray*}
The objective function of the pricing problem becomes $$(d+D\rho-\hat{\eta})^\top\Omega(\hat{\pi}-\hat{\eta}) = (D\rho)^\top\Omega(\hat{\pi}-\hat{\eta}) + const.$$ Constraints on $\rho$ must reflect that (i) $\bm{0} \leq \rho \leq \bm{1}$; (ii) if choice from one set is active, choice from all supersets thereof is active, (iii) if the default option is chosen from all subsets of a set, then it is chosen from the set as well.

In sum, the pricing problem can be expressed as the following integer linear program: 
\begin{eqnarray*}
\max_{\rho \in \{0,1\}^I} (D\rho)^\top\Omega(\hat{\pi}-\hat{\eta}) \\
\text{s.t.    }\rho_i-\rho_j &\geq & 0 \text{   whenever choice problem }i\text{ contains problem }j \label{eq:reg} \\
\rho_i &\leq & \sum_{j=1,\dots,k:x_j \in X_i} \rho_j.  \label{eq:reg2}
\end{eqnarray*}
\item[(iii)] At first glance, the tightened optimization problem \eqref{eq:J*N} has no sparse solution, but \citet{RUM_compute} remedy this. Heuristically, the vector $\bm{1}\cdot\tau_N/H$ can be concentrated out of the problem and a problem with sparse solution remains. A catch is that this requires the initial guess $\tilde{\boldsymbol{BA}}$ to have the same dimension as the true $A$ (its column cone cannot be contained in a face of the $\mathcal{C}$). \citet{RUM_compute} generate columns at random and verify that this constraint is met. This will not work here because only one of possibly millions of choice types makes a default choice on the universal set. Random column generation would be unlikely to discover that type, and so we seed $\tilde{\boldsymbol{BA}}$ with the corresponding column, $300$ additional random columns, and verify the rank condition. This is a problem and a fix that is likely to apply to other applications of the method as well.
\end{itemize}

\section{Supplementary Appendix: \newline Testing for Choice Overload Against the Benchmark of RSM}\label{app:rsm}

We now describe how to test if choice from $X$ exhibits choice overload with respect to random RSM given default choice probabilities $p_d:\mathcal{D} \rightarrow [0,1]$ on choice sets $\mathcal{D}$, $X \in \mathcal{D}$. Given a grand set $X$, there are a finite number of RSM models $(P_1,P_2)$. In principle, one could construct a $\boldsymbol{A}$ matrix with columns corresponding to the choices for each of the (finite) deterministic RSM choice patterns as we do for RUM in Section \ref{sec:population}. $\boldsymbol{B}$ can be constructed just as in Section \ref{sec:population} and $p_d^{RSM}(X)$ can be defined similarly to Corollary \ref{Corr1}. Our approach used for testing choice overload with respect to RUM can in principle be used to test for overload with respect to any model of choice that admits a finite number of deterministic choice types. 

However, with the 12 non-default alternatives of our experimental data, the number of admissible choice patterns is too large to compute $\boldsymbol{A}$.\footnote{$P_1$ makes $\boldsymbol{A}$ significantly larger than for RUM.} Instead, we find a characterization of $p_d^{RSM}(X)$ that works for any dataset $\mathcal{D}$ of the form of our experimental data. Given a set of alternatives $X$, let $\mathcal{B} = \{\{x,d\}: x \in X; \text{ } x \neq d\}$ and $\mathcal{T} = \{\{x,y,d\}: x,y \in X; \text{ } x,y \neq d\}$ be the sets of all binary and trinary subsets containing the default. As in our experiment, we assume $p_d$ is observed on sets $\mathcal{B} \cup \mathcal{T} \cup \{X\}$. Under this restriction, the number of observably distinct RSM types can be reduced. We say an RSM is $(P_1,P_2)$ is \emph{simple} if: (a) there is no $x$ for which $x P_1 d$ or $d P_1 x$ (the default is unblocked and does not block anything) (b) if $x P_1 y$, then $y P_2 d P_2 x$ (the only blocking that occurs is from alternatives worse than the default to those better). Admissible RSM choice patterns on the binary and trinary sets can be characterized by only considering simple RSMs. 

\begin{lemma}\label{lemma:simple_rsm}
If probabilities $p_d$ are consistent with random RSM on choice sets $\mathcal{B} \cup \mathcal{T}$, there exists a representation $\rho$ with $\rho(P_1,P_2)>0$ only if $(P_1,P_2)$ is simple.
\begin{proof}
    First we show show it is without loss to consider $d$ unblocked. Consider $(P_1,P_2)$ with $d$ blocked by some $x$: $x P_1 d$. This only affects choice from sets containing $x$, and as $P_1$ is acyclic guarantees active choice from all these sets. Deleting this blocking  ($x \not P_1 d$), setting $x P_2 d$, and deleting all blockings of $x$ ($z \not P_1 d$ for all $z \in X$) only can affect choices from these sets containing $x$ and also guarantees active choice from them. Repeating this procedure, we can modify $(P_1,P_2)$ to delete all blockings of $d$ without changing active / passive behavior on all small sets.

    We prove that it is without loss to consider RSMs without $d$ blocking anything. Take RSM $(P_1,P_2)$ with $d$ unblocked (as per the paragraph above) and suppose $d P_1 y$ for some $y \neq d$. Then $y$ is never chosen from any set. Removing this blocking and setting $d P_2 y$ guarantees the same choice is made from every set including $y$. 

    We now prove all blockings are of the form stated in the lemma. Take any RSM $(P_1,P_2)$ with $d$ unblocked and not blocking anything. In terms of behavior from sets $\mathcal{B} \cup \mathcal{T}$, a blocking between two non-default elements, $x P_1 y$, only affects choice from set $\{x,y,d\}$. If $x$ is better than $d$, there is active choice from this set ($P_1$ is acyclic so $y$ cannot block $x$) and removing the blocking does not change that. If $x$ and $y$ are both worse than $d$, then there is default choice from this set regardless of whether $x$ blocks $y$.
\end{proof}
\end{lemma}

We call a random RSM simple if it only has support on simple RSMs. For any deterministic RSM type $(P_1,P_2)$, the proof constructs a simple RSM that generates the same active/passive choice behavior on these sets. Finding $p_d^{RSM}(X)$ hence entails maximizing the probability of default choice from $X$ among simple random RSMs consistent with small set data. To do this, we note that many simple RSMs can be modified to choose $d$ from $X$ while making identical active/passive choice on the small sets. The following lemma states that the simple RSMs that \emph{cannot}, are the set rational utility maximizing types ($P_1$ empty) which prefer something to the default. Let $\mathcal{P}^*$ be this set of types: $\mathcal{P}^* = \{(P_1,P_2): P_1 \text{ empty; } xP_2d \text{ for some } x \neq d \}$. Every type in $\mathcal{P}^*$ must make active choice from $X$. Furthermore, for every RSM type in $\mathcal{P}^*$, there is no RSM type that makes identical choices on binary and trinary sets and chooses $d$ from $X$. The converse is true for all simple RSM types $(P_1,P_2) \not \in \mathcal{P}^*$: for every such type, either $(P_1,P_2)$ picks $d$ from $X$ or there is another RSM type $(P_1',P_2')$ which makes identical choices on $\mathcal{B} \cup \mathcal{T}$ and chooses $d$ from $X$. Then:

\begin{lemma}\label{lemma:active_rsm}
There is no RSM type which makes identical active/passive choices to simple RSM type $(P_1,P_2)$ on $\mathcal{B} \cup \mathcal{T}$ and chooses $d$ from $X$ if and only if $(P_1,P_2) \in \mathcal{P}^*$.
\begin{proof}
`Only if'. Take any simple RSM $(P_1,P_2)$ for which (b) does not hold: then $d$ is chosen from all sets. Take any simple RSM for which (a) does not hold; there exists some $x,y$ with $xP_2dP_2y$ and $y P_1 x$. Modify $P_1$ to add the following blockings: $x P_1 z$ for all $z \neq x$ with $z P_2 d$ (i.e. $x$ blocks everything else better than the default); $P_1$ remains acyclic as the original RSM was simple. This new RSM makes the same choice from all small sets ($x$ blocking another alternative $z$ better than the default still guarantees active choice from $\{x,z,d\}$). It chooses $d$ from $X$ as all alternatives better than $d$ are blocked by something.

`If'. Take any $(P_1,P_2)$ satisfying (a) and (b); let $\bar X$ be the set of alternatives better than default under $P_2$. For any (potentially nonsimple) RSM $P'=(P_1',P_2')$ to agree with $(P_1,P_2)$ on sets $\mathcal{B}$ and pick $d$ from $X$ it must have all elements in $\bar X$: (1) better than $d$ and unblocked by $d$ (we need active choice from all $\{x,d\}$ for $x \in \bar X$; we cannot block $d$ as then we have active choice from $X$), and (2) all these elements blocked by something. Then there must be some $x \in \bar X$ that is blocked by some $y \not \in \bar X$, $y \neq d$ (by acyclicality of $P_1'$, all elements of $\bar X$ cannot block one another). But then we have default choice from set $\{x,y,d\}$ under $P'$ while we have active choice under $(P_1,P_2)$.

\end{proof}
\end{lemma}

Our main result in this section follows immediately from Lemma \ref{lemma:active_rsm}. Given a simple random RSM $\rho$ consistent with the data, we can find another consistent random RSM such that all types outside $\mathcal{P}^*$ in its support pick $d$ from $X$. Meanwhile, we cannot modify types in $\mathcal{P}^*$ to choose $d$ from $X$. Hence,

\begin{theorem}\label{thm:rsm}
Suppose $\mathcal{D} =\mathcal{B} \cup \mathcal{T} \cup \{X\}$. For any $p \in [0,1]$, $p_d^{RSM}(X) \geq p$ if and only if there exists a simple random RSM $\rho$ with $\rho(\mathcal{P}^*) \leq 1-p$.
\end{theorem}

$p_d^{RSM}(X)$ can be found explicitly by minimizing the probability assigned to types $\mathcal{P}^*$ across all simple random RSMs that explain the binary and trinary set data, or equivalent maximizing the probability assigned to the complement of $\mathcal{P}^*$:  $$p_d^{RSM}(X) = \max_{\{\rho \text{ simple, consistent with data on }\mathcal{B} \cup \mathcal{T}\}} 1- \rho(\mathcal{P}^{**}).$$

However, computing this value directly requires constructing an $\boldsymbol{A}$ matrix with columns pertaining to every simple RSM type. While this is significantly smaller than the full of set of RSM types, it is still too large with $|X|=12$ to practically compute this matrix.\footnote{Simple RSM types can be fully described by the set of alternatives that are better than the default under $P_2$, and the set of blocking pairs $xP_1y$ where $yP_2dP_2x$. There are hence (on the order of) $2^{12}$ relevant $P_2$ orderings (the same as under RUM), but for each of these we have to consider all $P_1$'s under the simple RSM restriction.} We instead use the theorem to lower bound $p_d^{RSM}(X)$ and prove that $p_d^{RSM}(X) > p_d(X)$ in our data and hence that we do not see choice overload with respect to random RSM.\footnote{The approach is similar to \cite{RUM_compute}'s approach for testing RUM when the number of types is too large to compute $\mathbf{A}$.}

Given any subset of simple RSM types $\mathcal{P}'$ which includes all RUM types (and hence all of $\mathcal{P}^{*}$), we can compute the a A-matrix for just these types, $\mathbf{A}'$. This is computable if we add a small enough number of simple RSM types to the rational types in the RUM matrix. Let $\hat \pi$ be the vector empirical choice probabilities from the binary and trinary sets as in in Section \ref{sec:population}. Suppose there is a solution to the equation from Corollary \ref{Corr1}, $\mathbf{B} \mathbf{A}' \nu = \hat \pi$, so the small set data is consistent with a random RSM representation using only types $\mathcal{P}'$. Let  $\nu^*$ be the representation which minimizes the probability assigned to types $\mathcal{P}^{*}$, and let $q$ be this probability; then we know $p_d^{RSM}(X) \geq 1-q$. Adding more non-rational RSM types to $\mathcal{P}'$ reduces $q$ (weakly), as we have more types outside of $\mathcal{P}^*$ which we can use (in particular, it is smallest with the matrix including all simple RSM types). If for any $\mathcal{P}'$ we find that $p_d(X)<1-q$, we can conclude there is no choice overload in our data. 

The types violations of RUM that drove choice overload our data allowed us to systematically find $\mathcal{P}'$. In particular, the main violations we see pertain to the two `good' alternatives which yield more experimental points than the default, $\{9,11\}$. We see high active choice from the binary sets with these alternatives than we see from some trinary sets, e.g. $\{x,9,d\}$, $\{x,11,d\}$. We added to the A-matrix simple RSM types with $9$ and/or $11$ blocked by different subsets of other alternatives (as well as some other similar types).\footnote{The complete set of simple non-RUM RSM types added were all $(P_1,P_2)$ in the following classes: (1) $xP_2d$ for exactly one $x \in X \setminus \{d\}$, and a set of blockers $B \subset X \setminus \{d,x\}$ with $y P_1 x$ for $y \in B$; 804 such types. (2) $9 P_2 d$ and $11 P_2 d$ with all other alternatives worse than $d$, and each of $9$ and $11$ blocked by exactly one alternative from $X \setminus \{d,9,11\}$ or by all alternatives $X \setminus \{d,9,11\}$; 121 such types. (3) $xyP_2d$ for some $x,y \in X \setminus \{d\}$, and $x$ and $y$ blocked by all other non-default alternatives; 66 such types.} In total, we added 991 non-rational RSM types, addition to the 4096 rational types. We did so until applying the \cite{KS18} test from Section \ref{sec:RUM_test} to the resulting $\mathbf{A}'$ matrix did not reject that there is a solution to  $\mathbf{B} \mathbf{A}' \nu = \hat \pi$, meaning our data is (accounting for sampling noise) consistent with a random RSM representation using only these types. Minimizing the probability assigned to types $\mathcal{P}^*$, we obtain a bound of $p_d^{RSM}(X) \geq 0.872$, which is far above the default choice we see on our data, $p_d(X) = 0.223$. 

We conclude by clarifying tightness of the bound in Example \ref{ex:rsm}.

\begin{lemma} \label{lemma:rsm_example}
In Example \ref{ex:rsm}, we have $p_d^{RSM}\leq 3/4$.
\end{lemma}
\begin{proof}
 If an RSM model predicts default choice in $X$, then it must also predict choice of default in some trinary set. This is because, if the default is chosen in $X$, then it must not be blocked by any element in $X$, and every element that is better than it must be blocked by some other element. If there are more than two elements worse that $d$, then $d$ must be chosen from the trinary set that also contains those two elements. If not, given that $P_1$ is acyclic, it cannot be the case that all elements that are better than $d$ are blocked by other elements that are better than $d$: at least one must be blocked by the element worse than $d$, in which case the set with these two items in will also lead to the choice of $d$. Thus the probability choice of default in $X$ must be no higher than the sum of default choices across such sets.
 \end{proof}

\end{document}